\documentclass[fleqn,12pt]{SelfArx} 

\usepackage[english]{babel} 
\usepackage{tikz}
\usepackage{natbib}
\usepackage{amsmath, amstext}
\usepackage{changepage}
{\setlength{\mathindent}{4mm}

\setlength{\columnsep}{0.55cm} 
\setlength{\fboxrule}{0.75pt} 

\definecolor{color1}{RGB}{0,0,90} 
\definecolor{color2}{RGB}{0,20,20} 

\usepackage{hyperref} 
\urlstyle{same}
\hypersetup{hidelinks,colorlinks,breaklinks=true,urlcolor=color2,citecolor=color1,linkcolor=color1,bookmarksopen=false,pdftitle={Title},pdfauthor={Author}}

\JournalInfo{Robotic Telescopes, Student Research and Education (RTSRE) Proceedings} 
\Archive{ Conference Proceedings, San Diego, California, USA, Jun 18-21, 2017 \\Fitzgerald, M., James, C.R., Buxner, S., White, S., Eds. Vol. 1, No. 1, (2018) \\ ISSN XXXX-XXXX (online) / doi : doidoidoidoi / CC BY-NC-ND license \\ Peer Reviewed Article. rtsre.org/ojs} 
\PaperTitle{FTP and NSO: Astronomy With (Several) Robotic Telescopes} 

\Authors{Fraser Lewis\textsuperscript{1, 2}*} 
\affiliation{\textsuperscript{1}\textit{Faulkes Telescope Project, School of Physics \& Astronomy, Cardiff University, The Parade, CF24 3AA, Cardiff, Wales}}
\affiliation{\textsuperscript{2}\textit{Astrophysics Research Institute, Liverpool John Moores University, 146 Brownlow Hill, Liverpool L3 5RF, UK}} 

\affiliation{*\textbf{Corresponding author}: fraser.lewis68@gmail.com} 

\newcommand{\keywordname}{Keywords} 

\Abstract{We present details of two UK based robotic telescope projects, The Faulkes Telescope Project and the National Schools Observatory. We discuss details on how these projects utilise large aperture robotic telescopes for education purposes.}

\begin{document}

\maketitle

\begin{tikzpicture}[remember picture,overlay]
   \node[anchor=north west,inner sep=10pt, xshift=1.5cm, yshift=-0.25cm] at (current page.north west)
              {\includegraphics[scale=2.0]{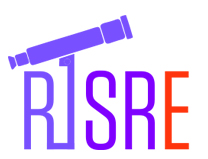}};
\end{tikzpicture}

\flushbottom 

\maketitle 


\thispagestyle{empty} 

\section*{Introduction}

\addcontentsline{toc}{section}{Introduction}

\begin{flushleft}

The Faulkes Telescope Project (FTP) and the National Schools Observatory (NSO) are two of the earliest robotic telescope projects that brought truly research grade metre+ class telescopes online for education use \citep{newsam2003liverpool}. Earlier projects tended to use either relatively small aperture amateur grade telescopes or re-purposed older instrumentation such as the 30” Leuschner Telescope or the 24” Mt Wilson Telescope \citep{gomez2017robotic}. In this article, the history of both the FTP and the NSO is presented as well as their current status and future directions.\\

\end{flushleft}


\section*{The Faulkes Telescope Project}

\addcontentsline{toc}{section}{The Faulkes Telescope Project}
\begin{flushleft}

Based in Cardiff, Wales since 2004, The Faulkes Telescope Project (FTP) (\citealt{roche2005faulkes}, \citealt{rochep2008}, \citealt{lewis2010robotic}) was originally conceived by Dr Martin ‘Dill’ Faulkes as a way of promoting the teaching of STEM subjects (especially though the medium of astronomy) in schools in UK and Ireland. Two 2-metre aperture telescopes of Ritchey-Cr\'etien design were built by Telescope Technologies Ltd (TTL), a spin-off company of Liverpool John Moores University (LJMU). These two f/10 robotic telescopes, shown in Figure \ref{fig1}, are located at Haleakala on Maui, Hawai’i (FT North; FTN) and Siding Spring in New South Wales, Australia (FT South; FTS). In 2006, FTN and FTS were bought by the the Las Cumbres Observatory (LCO; as detailed in \citealt{gomez2018}, these proceedings, \citealt{brown2013cumbres}) and since then, FTP has been an educational partner of LCO.\\
\bigskip

\begin{figure*}\centering
\includegraphics[width=0.9\linewidth]{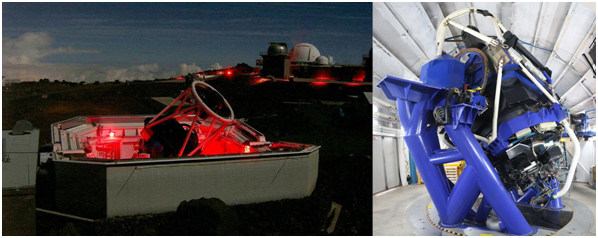}
\captionsetup{format=hang}
\caption{\small{\textit{The Faulkes Telescopes. FTN on the left, FTS on the right}}}
\label{fig1}
\end{figure*}

The LCO network now has a further nine 1-metre telescopes along with ten 0.4 metre telescopes. FTP provides free access (via the internet) to all of these telescopes via LCO’s Valhalla queue-scheduler and real-time access to two metre and 0.4 metre telescopes in both Hawai’i and Australia. This access is exclusively for educational users, predominantly in the UK and Ireland but also including educators and schools from Europe and the USA.\\
\bigskip

Approximately 1000 hours are available per year to these users free-of-charge, allowing school students and teachers alike to use research grade instrumentation. They are able to access imaging cameras to create single filter or colour images as well as for more scientific projects. Access is also available to the FLOYDS low-resolution spectrographs available on the 2-metre telescopes. \\
\bigskip

In addition to providing telescope time, FTP provides resources on its website giving users information on what targets are suitable to observe, when they are visible, how to analyse the data (e.g. colour imaging, photometry)\footnote{\url{http://resources.faulkes-telescope.com/}} and access the data archive. Each month, a list of suitable and interesting targets are provided\footnote{\url{http://www.faulkes-telescope.com/targets}} allowing less experienced users the opportunity to use the telescopes without requiring a huge investment of time that teachers very often do not have available to them. In October 2017, the FTP welcomed its 1000th unique user.\\
\bigskip

Schools are encouraged to take part in scientific research projects, using both archival and ‘pre-packaged’ datasets (for an example, see the work of Bartlett in these proceedings). These projects include the detection of asteroids, light-curves of eclipsing binaries, pulsating variables (such as Cepheids, delta Scuti and RR Lyrae stars), transiting exoplanets and the creation of Colour-Magnitude Diagrams (CMD). These CMD are fundamental diagrams in astronomy, allowing us to measure key parameters of populations of stars in open clusters, globular clusters and galaxies.\\
\bigskip

\begin{figure*}[ht]\centering
\includegraphics[width=0.89\linewidth]{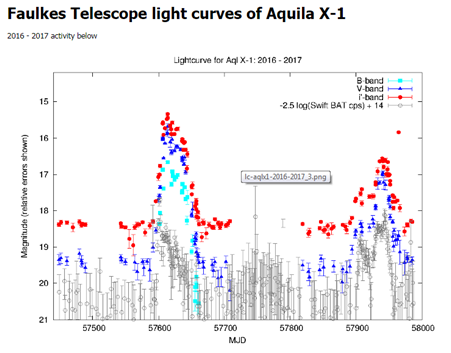}
\captionsetup{format=hang}
\caption{\small{\textit{An example light curve of the neutron star X-Ray Binary, Aquila X-1 (data reduced by David Russell, NYUAD with observations from schools and undergraduate students at NYUAD and Cardiff University)}}}
\label{fig2}
\end{figure*}

A key part of these projects is the ability to collaborate with other schools, with university researchers such as those at The Open University (see \cite{kolb2018}, these proceedings) and within other projects such as the EU FP7 Go-Lab Project\footnote{\url{http://www.golabz.eu/}} and the Erasmus + PLATON\footnote{\url{http://platon-project.eu/}} project. On occasion, the publication of these results\footnote{\url{http://www.faulkes-telescope.com/showcase/research}} is able to acknowledge the contribution made by FT users (e.g. \citealt{elebert2009optical}, \citealt{lewis2010double}) in scientific journals.\\
\bigskip

In conjunction with Portuguese collaborators at NUCLIO\footnote{\url{http://nuclio.org/}} and researchers at New York University Abu Dhabi (NYUAD), we have developed a research and education project (see Figure \ref{fig2}) entitled Black Holes In My School\footnote{\url{http://www.golabz.eu/spaces/black-holes-my-school}}\textsuperscript{,}\footnote{\url{http://www.faulkes-telescope.com/XRB}}. Further, undergraduate students at NYUAD have produced Astronomers Telegrams (ATels) and contributed to papers as authors having been involved not just in the collection of data but in more advanced analysis such as the photometry and interpretation of results (e.g. \citealt{qasim2016follow}, \citealt{qasim2017optical} and \citealt{russell2018optical}).\\
\bigskip

As detailed by \citeauthor{bartlett2018} (2018, these proceedings), the FTP also encourages its users to contribute to the Gaia Alerts program\footnote{\url{http://gsaweb.ast.cam.ac.uk/alerts/alertsindex}}. Of particular interest are the supernovae detected by ESA’s Gaia satellite \citep{brown2016gaia} and especially, the Type Ia supernovae. Their link with cosmology and the measurement of the expansion of the Universe is one of only a few common threads in science curricula worldwide.\\

\end{flushleft}

\section*{The National Schools’ Observatory}

\addcontentsline{toc}{section}{The National Schools’ Observatory}

\begin{flushleft}

Complementary to the work of the FTP is the work of the The National Schools’ Observatory (NSO)\footnote{\url{http://www.schoolsobservatory.org.uk/}} \citep{newsam2007national}, which uses time on the 2-metre Liverpool Telescope (LT) (\citealt{steele2000enabling}, \citealt{steele2004liverpool}). The LT is based at the Instituto de Astrofísica de Canarias (IAC) at Observatorio del Roque de los Muchachos on La Palma in the Canary Islands, Spain. Also built by TTL, the LT features a broader range of instrumentation than FTN/FTS and is run by the Astrophysics Research Institute (ARI) at Liverpool John Moores University.\\
\bigskip

\begin{figure*}[ht!]\centering
\includegraphics[width=0.9\linewidth]{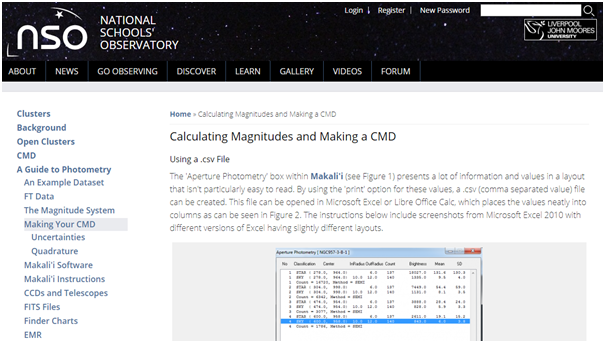}
\captionsetup{format=hang}
\caption{\small{\textit{Example screenshot of a page from the NSO open clusters activity}}}
\label{fig3}
\end{figure*}

Established in 2004, the NSO provides schools in the UK and Ireland with access to the Liverpool Telescope through a guided observing system, using 10\% of the LT’s time. Its website contains astronomy related content, news and learning activities. Already with over 4,000 users, new in 2017 is the ability for non-UK/Ireland based schools and teachers to register affording free access to both the observing portal and resources.\\
\bigskip

Amongst this material (and designed by the author) is an inquiry-based ‘teacher-free’ activity for students to learn about open clusters and HR diagrams as well as the technique of photometry\footnote{\url{http://www.schoolsobservatory.org.uk/discover/projects/clusters/main}}. Including screenshots and screencasts, this activity shows students how to analyse their data (using a spreadsheet package such as Excel) and to upload their results in the form of a CMD (see Figure \ref{fig3}) with the aim of encouraging them to discuss their findings with other students.\\
\bigskip

Currently, there is a choice of 28 datasets (B and V images from the FTP) but teachers may wish to take their own observations of a cluster of their choice with FTP or NSO. There is even the opportunity here to use the FLOYDS (FTN/FTS) or SPRAT (LT) spectographs to follow-up any object of particular interest, such as extremely blue or red stars. This activity is the first of several projects which will be created in coming months on topics such as exoplanets and variable stars.\\

\end{flushleft}

\section*{Future Directions}

\addcontentsline{toc}{section}{Future Directions}

\begin{flushleft}

For the future, FT continues to work with schools and teachers worldwide strengthening links with schools from Europe and further afield by continuing to provide access to LCO’s expanding global network. The NSO looks forward to the likely development of a 4-metre telescope owned by LJMU and located next to the Liverpool Telescope. It is envisaged that this facility will be running alongside the existing LT providing even more opportunities for researchers and educators alike.\\

\end{flushleft}


\bibliographystyle{apalike}

\bibliography{references}

\end{document}